\documentclass{nature}
\usepackage{hyperref,longtable}
\usepackage{graphicx}

\title{A Practical Approach to Spatiotemporal Data Compression}

\author{Niall H. Robinson$^{1}$, Rachel Prudden$^1$ \& Alberto Arribas$^1$}

\begin{document}

\maketitle

\begin{affiliations}
 \item Informatics Lab, Met Office, Exeter, UK.
\end{affiliations}

\begin{abstract}
Datasets representing the world around us are becoming ever more unwieldy as data volumes
grow. This is largely due to increased measurement and modelling resolution, but the problem is often exacerbated when data are stored at spuriously high precisions. In
an effort to facilitate analysis of these datasets, computationally
intensive calculations are increasingly being performed on specialised
remote servers before the reduced data are transferred to the consumer.
Due to bandwidth limitations, this often means data are displayed as
simple 2D data visualisations, such as scatter plots or images. We
present here a novel way to efficiently encode and transmit 4D data
fields on-demand so that they can be locally visualised and
interrogated. This nascent ``4D video'' format allows us to more
flexibly move the boundary between data server and consumer client.
However, it has applications beyond purely scientific visualisation, in
the transmission of data to virtual and augmented reality.
\end{abstract}

With the rise of high resolution environmental measurements and
simulation, extremely large scientific datasets are becoming
increasingly ubiquitous. The scientific community is in the process of
learning how to efficiently make use of these unwieldy datasets.
Increasingly, people are interacting with this data via relatively thin
clients, with data analysis and storage being managed by a remote
server. The web browser is emerging as a useful interface which allows
intensive operations to be performed on a remote bespoke analysis
server, but with the resultant information visualised and interrogated
locally on the client\cite{Shen2014, Jupyter}.

There is a also a widespread desire to allow the public better access to
data. Indeed, this is now often a stipulation of taxpayer funded
research. Mere availability of the raw data is no longer considered
satisfactory, and researchers are often asked to give more practical
access to
information\cite{Hubbard2014, unleashing}.
The web browser is the natural portal for the public to consume this
data.

Many of these large datasets are highly multidimensional, and often
spatiotemporal. For instance, the field of earth science is generating
extremely large spatiotemporal datasets on a daily basis, from weather
forecasts to climate simulations: the UK's Met Office will soon be generating
$\sim$400\,TB daily, and thier our archive is approaching 1\,EB.
Modern medical imaging also generates high resolution spatiotemporal
datasets from
scans\cite{Uecker2010}.

Compression algorithms traditionally used for images have previously
been applied to atmospheric
data\cite{Becker2015, Hubbe2013, Lucero2003} and medical
data\cite{Kim2009}.
The data from each time-step is first converted from a 3D to a 2D raster
grid by tiling slices adjacently, before encoding. Hubbe et al. 2013\cite{Hubbe2013}
concluded that lossy compression of climate data could be more widely
utilised.
While such image codecs are good at compressing data with spatial
coherence, they neglect the gains that can be made by compressing
temporal coherence.

Addressing ``data overload'' is one of the biggest challenges in modern
science. The work presented here investigates how we can empower data
consumers to interact with remote datasets. More specifically, we
present the first implementation of a practical and efficient method for
the dissemination of large spatiotemporal datasets with a focus on
compatibility with web technology by utilising video codecs.

The UK's Met Office generates world-leading weather forecasts several
times an hour. The raw forecast data is stored in a custom
meteorological format called GRIdded Binary version 2, or
GRIB2\cite{WMO}.
This format can implement extensive lossy or lossless compression of the
data, including optional implementation of PNG and JPEG 2000 codecs.
However, in practice, the operational lossy compression is often limited
to ``bit shaving''
- a technique which simply limits the precision of each datum. A
standard forecast field, consisting of latitude $\times$ longitude $\times$ altitude $\times$
time points, results in a $\sim$5\,GB GRIB2 file.

We limited the precision of all the data to 8-bit integers, that is, all
values are scaled to be between 0 and 255, which immediately quarters
the data volume. For each time point, the spatial 3D array was then
broken down into 2D latitude $\times$ longitude slices for each altitude level.
These slices were tiled adjacently and encoded as an image, with the
data points represented by the image pixel colours (Figure 1.). The
first third of the tiled slices were encoded in the red image channel,
the second third in the green channel, and the final third in the blue
channel. These images were then joined together into a
video.

\begin{figure}
\includegraphics[scale=0.8]{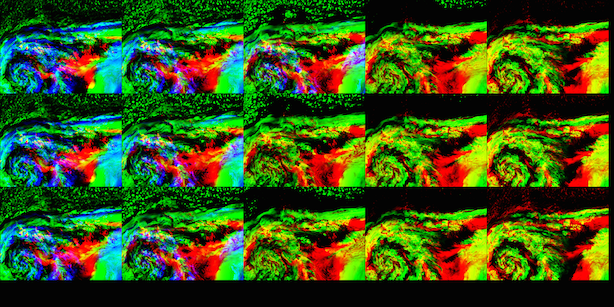}
\caption{3D atmospheric data of cloud
fraction encoded as pixels in an image.}
\end{figure}

We tested several widely available video compression algorithms which
give different levels of data compression and information loss, assessed
as datasets volume and Mean Absolute Error (MAE) respectively (Table 1). A test
dataset of forecast cloud fraction (i.e. $0.0 < x < 1.0
$) for the 27th November 2015 was used. Note that the major loss of
information occurs from the zlib compression during PNG encoding. The
various video compressions then have a negligible effect on information
quality, whilst drastically reducing the data volume.

\begin{longtable}[]{@{}lcc@{}}
\hline
& ~~~data volume~~~ & ~~~M.A.E. w.r.t GRIB2~~~\tabularnewline
\hline
\endhead
\textbf{GRIB2} & 5.0 Gb & n/a\tabularnewline
\textbf{8-bit GRIB2} & 1.25 Gb & 8.09e-4\tabularnewline
\textbf{8-bit pngs (zlib 6)} & 344 Mb & 0.162\tabularnewline
\textbf{MP4 x264} & 274 Mb & 0.163\tabularnewline
\textbf{Ogg Vorbis (q10)} & 128 Mb & 0.163\tabularnewline
\textbf{Ogg Vorbis (q2)} & 17 Mb & 0.166\tabularnewline
\hline
\caption{Data volume and information loss under different encodings.}
\end{longtable}

We then endeavoured to visualise these fields at a location remote to
the data via a web browser. A system was implemented to automatically
convert the forecast data to video. The process was resolved into
several microservices, written in Python. These microservices were
deployed using Docker Containers, making
them robust and portable. The whole process is automatically
orchestrated and executed in a compute cloud using Amazon Web Services.

For our prototype system, we chose to use the Theora Ogg Vorbis (q2)
codec, as it provides a good
compromise between compression ratio and compression speed. We also
chose Theora as it is open source, meaning is has the potential to be
extended to natively support 3D data in the future. The video compressed
version of the data is 10-20\,MB, which is a compression ratio of around
400:1 when compared to the original GRIB2 file.

\begin{figure}
\includegraphics[scale=0.4]{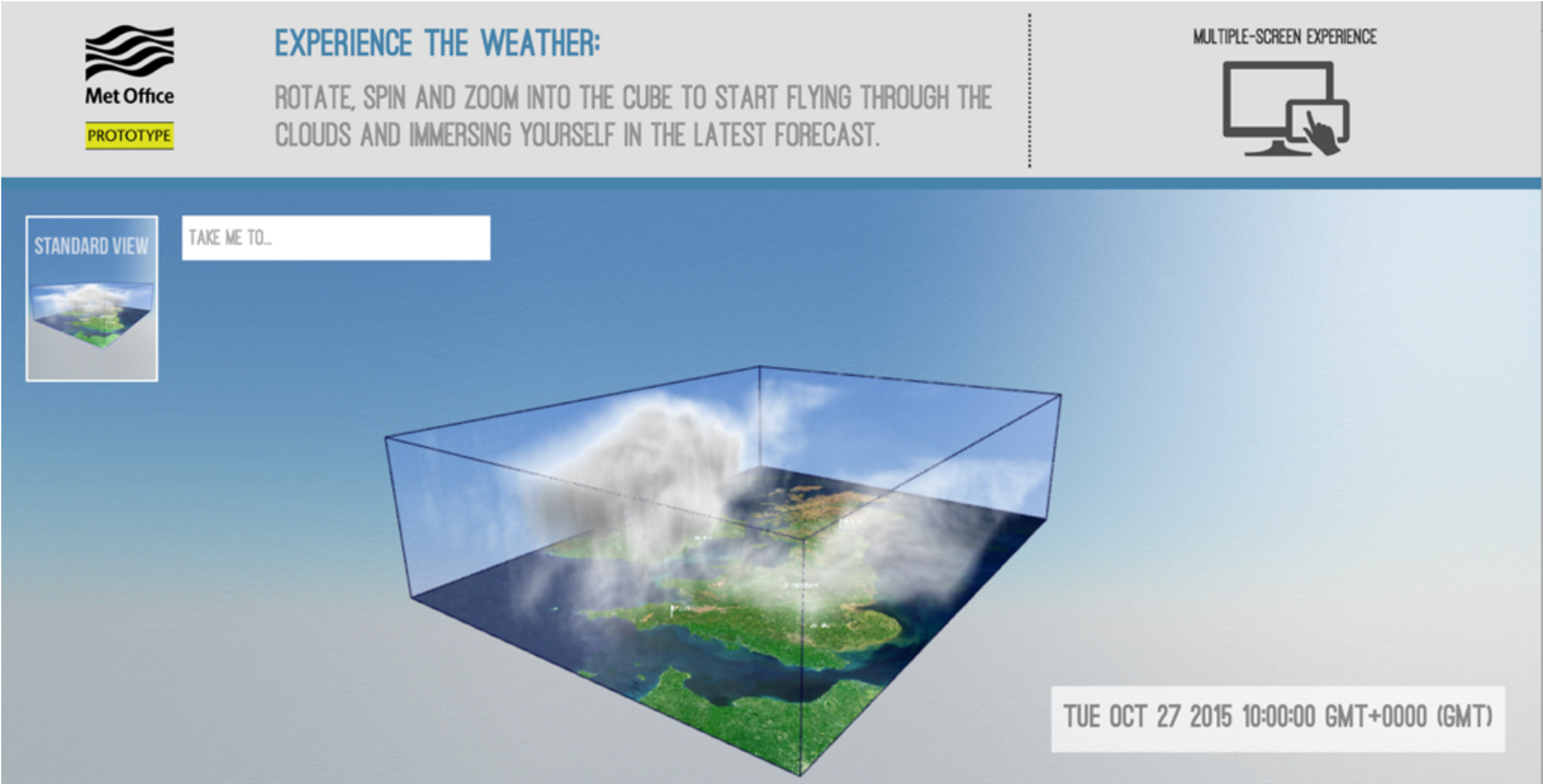}
\caption{The web application rendering the video encoded data as a 3D
field.}
\end{figure}

This video of atmospheric data is then served up to our web application
for rendering (Figure 2.). We created an
\href{http://demo.3dvis.informaticslab.co.uk/ng-3d-vis/apps/desktop/}{interactive
3D animation} of the data using
\href{https://www.khronos.org/webgl/}{WebGL} and bespoke
\href{https://www.opengl.org/documentation/glsl/}{GL Shader Language}
graphics card routines which simulated the passage of light rays through
the data (a technique known as volume rendering or ray tracing). This
application allows the users to interact with the animated 3D data field
over a standard internet connection, without installing specialist
software or hardware.

We set ourselves the task of representing weather forecasts in a way
that reflects all the generated data. This data is richly
spatiotemporal, however it is routinely communicated to the public as a
2D map, and scientists are largely limited to visualising data via
static 2D maps or 1D scatter plots. We wanted to implement our animated
3D visualisation in the web browser, both to make it widely accessible,
and to explore technologies which may eventually be of use to scientists
on web browser based thin clients. Encoding the data using video codecs
was central to achieving this.

Firstly, the use of video compression allowed us to significantly reduce
the data load. The 400:1 reduction in data volume is due to the loss of
data. Crucially though, the relevant information is retained: the
salient features of the data field are still present in the final
visualisation. Visual codecs are optimised to lose data which cannot be
seen, a feature which is not just optimal for traditional 2D
visualisation, but also largely suitable for such 3D rendering. More
generally, large datasets are routinely being stored at a spurious
precision, which is far higher than the physical precision of the model
or measurement. We think it is imperative that modern data scientists
examine their large datasets afresh, and consider new lossy approaches
to storing data at more appropriate and practical precision.

Employing video encoding is particularly useful in the context of
delivery to web browsers. They natively support the decoding of video
data, as opposed to esoteric atmospheric data formats. The video can be
easily streamed into the browser, meaning client memory is used
efficiently. As the dataset is represented graphically, it can easily be
transferred to the graphics card, where it can be rendered on the fly as
the user interacts with the visualisation. Other video functionality is
also useful, such as playback controls and on-the-fly scaling.

Over the past decade, the streaming of video content to web browsers has
been highly optimised by corporations such as YouTube and Netflix. We
propose that we must now consider similar approaches for the
transmission of ``3D video'', that is time dependent 3D rasters of data.

Currently, data can only be moved to client machines when it has been
reduced far enough. Efficient on-demand transfer of multidimensional
data will allow us to flexibly move the boundary between specialised
remote data servers for processing big datasets, and local client
machines for interrogation, visualisation and understanding. This
flexibility is essential to allow users (be they analysts or members of
the public) to fluently interact with data.

Simulations and measurements of the environment we live in seem set to
increase in application as well as volume. As virtual and augmented
reality technologies gain traction (both in consumer entertainment and
data analysis), it is imperative that we can broadcast dynamic content
for users. ``3D video'' should be allowed to become a fundamental and
common type of data, unhampered by limitations in dissemination.

The approach presented here, whilst far more optimised than previous
alternatives, can be built upon. Firstly, there is coherence in the
third spatial dimension which is currently not being leveraged by the compression algorithms. It is
also conceivable that an approach could be developed which is general
for n-dimensions, allowing efficient compression of highly
multidimensional datasets. Finally, video codecs are optimised to
preserve visual information, but work could be done to preserve more
esoteric properties, for instance, atmospheric turbulence information.

We have presented a novel but pragmatic approach to efficiently
disseminate 3D time dependent data to web browsers using video codecs.
Whilst our approach is simple, it has addressed a emerging fundamental
question: how can we communicate data which represents our environment?

\end{document}